\documentclass[10pt,a4paper,twoside]{aa}

\usepackage{epsfig,latexsym,longtable, graphicx}
\begin{document}


\def\sun{\hbox{$\odot$~}}
\def\deg{\hbox{$^\circ$}}
\def\kms{\,km\,s$^{-1}$}
\def\m{$^{\rm m}$}
\def\si{$\sim$}
\def\di{$\div$}
\def\av{A$_{\rm V}$ }
\def\msol{~M$_\odot$ }
\def\msolr{~M$_\odot$~yr$^{-1}$ }
\def\micron{\,$\mu$m}
\def\hi{H\,{\sc i} }
\def\marc{mag~arcsec$^{-2}$}
\def\cm2{cm$^{-2}$}
\def\ecs{ergs cm$^{-2}$ s$^{-1} \ $}
\def\es{ergs s$^{-1}$}
\def\cts{counts s$^{-1} \ $}
\def\esz{ergs s$^{-1}$ Hz$^{-1} \ $}
\def\kms{km s$^{-1}$}
\def\kcssk{$keV\ cm^{-2}\ s^{-1}\ sr^{-1}\ keV^{-1}\ $}
\def\ni {\noindent}
\def\Msun{$M_{\odot}\ $}
\def\lae{\mathrel{<\kern-1.0em\lower0.9ex\hbox{$\sim$}}}
\def\gae{\mathrel{>\kern-1.0em\lower0.9ex\hbox{$\sim$}}}

\title{The XMM-Newton and BeppoSAX view of the Ultra Luminous Infrared Galaxy MKN~231}

\author{V.\,Braito\inst{1,2}, R. Della Ceca\inst{1}, E. Piconcelli\inst{3,4},  P. \,Severgnini\inst{1}, L.\,Bassani\inst{3}, M.\,Cappi\inst{3},
A.\,Franceschini\inst{2}, K. Iwasawa\inst{5}, G.\,Malaguti\inst{3}, P. Marziani\inst{6},
G.G.C.\,Palumbo\inst{7}, M.\,Persic\inst{8},  G.\,Risaliti\inst{9,10} and  M.\,Salvati\inst{9}.}

\institute{
 INAF $-$ Osservatorio Astronomico di Brera, Milano, Italy
\and
Dipartimento di Astronomia, Universit\`a di  Padova, Italy
\and
IASF $-$ CNR, Bologna, Italy
\and
XMM-SOC (VILSPA),  ESA,  Madrid, Spain
\and
Institute of Astronomy, University of Cambridge, Madingley Road Cambridge CB3 OHA, U.K.
\and
 INAF $-$ Osservatorio Astronomico di Padova, Italy
 \and
Dipartimento di Astronomia, Universit\`a di Bologna,  Italy
\and
INAF $-$ Osservatorio Astronomico di Trieste,  Italy
\and
INAF $-$ Osservatorio Astrofisico di Arcetri, Firenze, Italy.
\and
 Harvard-Smithsonian Center for Astrophysics, Cambridge, USA}
\date{Received:XX; Accepted: XXX}
\authorrunning{Braito et al.}
\titlerunning{XMM-{\it Newton} and \emph{Beppo}SAX observations of MKN~231.}



\abstract{We discuss    XMM-{\it Newton}  and \emph{Beppo}SAX observations of MKN~231,  
the lowest-redshift  Broad  Absorption Line (BAL) QSO known so far and  one of the
best-studied Ultra Luminous  Infrared Galaxies.  By combining  the XMM-{\it
Newton} spectral resolution and the high-energy bandpass of \emph{Beppo}SAX  we
have been able to  study in more detail than previously  possible its
0.2--50 keV spectral properties.  The  \emph{Beppo}SAX  PDS data     
unveiled, for the first time,  a highly absorbed ($N_{\rm H}\sim 2\times
10^{24}$cm$^{-2}$) power-law component.  We find   that: a) below 10 keV we are
seeing only reprocessed radiation through reflection and/or scattering; b) the
intrinsic 2--10 keV luminosity of MKN~231 is $ 1^{+1.0}_{-0.5} \times
10^{44}$ ergs s$^{-1}$, i.e. more than an order of magnitude greater  than previous
measurements; c) the  starburst activity significantly contributes  to the  soft
($E<2$ keV) X-ray emission.
}

\titlerunning{XMM-{\it Newton} and \emph{Beppo}SAX observations of MKN~231}

\maketitle

 \section{INTRODUCTION}

In the local ($z < 0.05$) Universe Markarian ~231  ($z = 0.042$, $L_{bol}= 6 \times 10^{12}
L_{\odot}$ for H$_0=$ 50 km s$^{-1}$ Mpc$^{-1}$) is  one of the most luminous and  best studied Ultra
Luminous Infrared Galaxies (i.e. sources with $L_{IR}>10^{12} L_{\odot}$; hereafter ULIRGs). This
object  is optically classified as a  {\it Broad Absorption Line} (BAL)  QSO (Smith et al. 1995).
This class of sources, which  represent $\sim 10$\% of the optically selected QSOs, are known to be
weak X-ray sources, their faintness being attributed  to intrinsic absorption (Green et al.
\cite{gre01}; Gallagher et al. \cite{galla}).

Although the nature of the primary energy source (AGN vs.
starburst activity) for the high luminosity  of ULIRGs   is
still a debated issue (see Ptak et al. 2003, Franceschini et al.
2003 and references therein), in the case of MKN~231 optical and
near infrared observations seem to suggest that a   significant
contribution to the infrared luminosity ($L_{\rm{FIR}}\simeq 3\times 10^{12}L_{\odot}$) could be ascribed to AGN
activity (Goldader et al. \cite{golda}; Krabbe et al.
\cite{krab}).   On the other hand this galaxy is also
undergoing an energetic starburst and, in particular, radio
observations of the inner  kiloparsec regions  suggest  that   a
nuclear ring of active star formation is proceeding at a rate   of
$\sim 220\ M_{\odot}\ $yr$^{-1}$ (Taylor et al. 1999)\footnote { In
this respect it is worth  noting that  the starforming rate in
M82, the prototype local starburst galaxy, is about one order
of magnitude lower (Grimm et al. \cite{grim})}.   Indication of a
strong starburst activity, which can account for up to 70\% of the
$L_{FIR}$ of MKN~231  (corresponding to an SFR $\sim 470\pm50 \
M_{\odot}\ $yr$^{-1}$), comes also from the  modelling of the radio
to optical spectral energy distribution    (Farrah
et al. 2003). This  latter estimate is in good agreement with the
AGN fraction derived by Lonsdale  et al. (2003) using  VLBI
observations at 18 cm.  As discussed by several authors the combination
of starburst and luminous  AGN makes MKN 231 one of the best
examples of the transition from starburst  to AGN according to the
scheme outlined in Sanders et al. (\cite{san}).

The strong circumnuclear starburst activity together with the presence of
high intrinsic absorption,  probably associated with   the BAL outflows (Green et al. \cite{gre01}),
makes the X-ray continuum of this source particularly  complex. Many
different spectral components should be taken into account  in the spectral
modeling i.e. the thermal and the binary emission  associated
with the products of the starburst as well  as the scattered/reflected and intrinsic
component(s) associated with the  AGN.

Previous X-ray observations by ROSAT, ASCA  (Turner 1999, Iwasawa
1999; Maloney \& Reynolds 2000, hereafter MR00) and more recently
{\it Chandra} (Gallagher et al. 2002, hereafter G02; Ptak et al.
2003),  revealed the presence of: a)  extended soft X-ray
emission of thermal origin, likely associated with the
circumnuclear starburst; b) a hard and flat  (photon index
$\Gamma \sim 0.7$) non-thermal component having an observed  X-ray
luminosity L$_{(2-10)} \sim 10^{42} $ergs s$^{-1}$  and c) a Fe
$K_{\alpha}$ emission line having  an observed equivalent width of
$\sim$ 300 eV (the line was measured by ASCA but is not  detected
by {\it Chandra}).

To explain the low observed luminosity (which is at least two  orders of 
magnitude lower   than that expected for an AGN of similar bolometric
luminosity), the flat photon index  and the EW of the Fe $K{\alpha}$ line,
MR00 suggested that about 3/4 of the observed  2--10 keV spectrum of MKN 231 consists of a scattered
power-law component plus a reflected component (no direct AGN component is seen
below $\sim$10 keV). Both  reflected and  scattered  components
are absorbed by an intrinsic column density of $N_{\rm H}\sim 3\times 10^{22}$
cm$^{-2}$. On the contrary G02, using {\it Chandra} data (which in addition 
make it possible to disentangle  the nuclear from the extended  component),  support a
model in which a small, Compton-thick absorber blocks the direct X-rays. Only
indirect scattered emission from multiple lines of sight can reach the observer
in the {\it Chandra} energy range; in particular they  proposed  a triple
scattering model composed of three separate absorbed power laws having a photon
index fixed to 2.1 and absorbing column densities of  $N_{\rm H} \simeq 1.2\times
10^{21}$cm$^{-2}$, $N_{\rm H}\simeq 3.2\times 10^{22}$cm$^{-2}$ and  $N_{\rm H}\simeq
5.9\times 10^{23}$cm$^{-2}$, respectively. According to the modeling of
MR00 and G02, the X-ray emission  below 10 keV is only reprocessed
radiation, thus they could  evaluate the intrinsic
luminosity  of MKN 231 only through indirect arguments.\\

Here the results from XMM-{\it Newton} and \emph{Beppo}SAX observations of this object are presented.
The main result from these data is that a highly absorbed ($N_{\rm H}\sim 2\times 10^{24}$ cm$
^{-2}$) AGN component having an intrinsic 2--10 keV luminosity of $1^{+1.0}_{-0.5}\times 10^{44}$
ergs s$^{-1}$ has been detected. {\bf This is the first direct measurement of the intrinsic power of
the AGN hosted by MKN 231.} This paper is organized as follows: observation and data analysis are
described in Sect.~2; timing and spectral analysis are reported in  Sect.~3 and 4. Results and 
discussion  are presented in Sect.~5. In this paper we assume  $H_0= 50$ km s$^{-1}$ Mpc$^{-1}$ and
$q_0=0.5$.  

 \section {Observations and Data reduction}

MKN~231 is part of an XMM-{\it Newton} survey of 10 nearby (z $<$ 0.2)
ULIRGs  for which high-quality mid-IR and optical spectroscopic data are
available. Results on this project are presented and discussed in  Braito et al. (2002) and in
Franceschini et al. (2003). MKN~231 also
belongs to a small sample of ULIRGs for  which we have obtained \emph{Beppo}SAX
measurements.

\subsection{XMM-{\it Newton} data}

MKN~231 was targeted for 20 ks by XMM-{\it Newton} on July 6th, 2001. The
EPIC cameras (European Photon Imaging Camera:  \cite{stru} and
\cite{Turner}) were operating  in full-frame mode with the thin filter
applied. Event files produced from the standard pipeline  processing have
been analyzed and filtered (using  version 5.4 of the Science Analysis
Software, SAS, and  the latest calibration files released by the EPIC
team) from high background time intervals and only events corresponding
to pattern 0-12 for MOS and pattern 0-4 for pn have been used  (see the
XMM-{\it Newton} Users' Handbook, \cite{Ehle2001}).  The net exposure
times, after data cleaning, are $\sim$ 16.0 ksec, $\sim$ 19.9 ksec and
$\sim$ 19.8 ksec for pn, MOS1 and MOS2 respectively.

In the low-energy domain (E $<$ 2 keV), the X-ray emission of MKN 231 appears to be
extended\footnote{A detailed spectral and imaging analysis of the
extended component will be reported elsewhere.}, with  a roughly
circular symmetry. The
presence of  extended soft X-ray emission was first
 noted using  ROSAT observations, (Turner \cite{Turner99})  and
it has been confirmed by the recent {\it Chandra} observation
(G02). The {\it Chandra} data show that the galaxy emission (0.35--2 keV range)
extends   over a radius of $\sim 25 $
arcsec. The  hard (2--10 keV)  X-ray brightness profile is
comparable to the XMM-{\it Newton} PSF, whose HEW  at $\sim$ 8
keV is $\sim 14''$.

Since in this paper we want to address the X-ray spectral
properties of the nuclear and circumnuclear X-ray emission,
source counts have been extracted from a circular region of 20
arcsec radius\footnote{The  X-ray source detected by G02 using
{\it Chandra} data at 3$''$ apart from the nucleus is thus included in the counts extraction region.
However, as  
     its flux is two orders of magnitude lower  than that of the nuclear emission of MKN~231, it does not
affect the results reported here.}, positionally coincident with
the core of the hard X-ray emission. The X-ray position is also
coincident (within the XMM-{\it Newton} positional accuracy $\sim$
few arcsec) with the core of the  optical and radio emission.  
The XMM-{\it Newton} PSF does not allow us to select  a smaller
region and thus,  although  the  extraction radius has been chosen
to minimize the effect of the extended soft emission, in the
derived spectrum   we expect imprinted signatures of  the
strong circumnuclear  starburst activity of MKN~231.

 The presence of other sources around MKN~231 prevented the selection of a nearby    large
enough  area to accumulate background counts with good statistics; therefore background
counts have been accumulated using the background event files distributed by the EPIC  team. 
Response matrices (that include the correction for the effective area)  have been generated
using the SAS tasks {\it arfgen} and {\it rmfgen}. The net count rate in the 0.2--10 keV
energy range is  $(11 \pm  1)\times 10^{-2}$ cts/s,  $(3.77\pm 0.14)\times 10^{-2}$ cts/s 
and  $(3.54\pm 0.14)\times 10^{-2}$ cts/s  for pn, MOS1 and MOS2, respectively.

\subsection{ \emph{Beppo}SAX data}

MKN~231 was observed by  \emph{Beppo}SAX (Boella et al., \cite{boella}) from December
29th, 2001 to January 1st, 2002. In this paper we use data collected from
the Medium Energy Concentrator Spectrometer (MECS) and from the Phoswich
Detector System (PDS). We do not use data from the Low Energy Concentrator
Spectrometer  since they are of significantly lower quality when compared
with the  low energy XMM-{\it Newton} data.

Standard data reduction was performed  using the software package SAXDAS (see
http://www.sdc.asi.it/software and the Cookbook; Fiore, Guainazzi \& Grandi 1999) for the
MECS datasets and XAS (Chiappetti \& Dal Fiume 1997) for the PDS data. Spectral fits were
performed using public response matrices as from the 1998 November issue.  \emph{Beppo}SAX
pointed at MKN~231 for an  effective exposure time of 144 ksec and 76 ksec for MECS  and PDS,
respectively. To maximize statistics and signal-to-noise ratio, the MECS source counts  were  extracted from a circular region of 2 arcmin radius. The background spectrum was taken 
from blank sky fields at the position of the source with the same extraction radius.  As
suggested in the \emph{Beppo}SAX Cookbook we have considered only the  MECS data in the
1.8-10 keV  energy range.  Spectral channels corresponding to energies in the range   15--60
keV were used for the PDS. MECS source counts have been rebinned proportionally to the
instrumental resolution and so as  to have always more than 20 counts per channel; the PDS
data instead have been rebinned to maximize the signal to noise ratio in each energy bin. 
The net count rates are $(5.68\pm0.23)\times10^{-3}$ cts/s in the MECS and $(6.58 \pm 2.16)\times10^{-2}$ cts/s in the PDS.

\section{X-ray variability}

The {\it Chandra} data of MKN 231 discussed by G02 indicate significant
variability with a count rate decrease of $\sim 45$\%
on a 20 ks time-scale at energies above 2 keV. This variability coupled with the 188
eV upper limit on the EW of the  Fe $K{\alpha}$ emission line was used
by G02 to argue  against the reflection model
proposed by MR00.

Thus we  first investigated the variability properties of MKN
231  using the {\it Beppo}SAX data. Although \emph{Beppo}SAX  has
a significantly worse spatial resolution than {\it Chandra}, the long exposure
time (144 ks) of this pointing is ideal to look for variability on
similar time scales as those investigated by G02. Using the
\emph{Beppo}SAX MECS counts in the 2--10 keV energy range
\footnote {The \emph{Beppo}SAX PSF does not allow us to extract a
nuclear light curve, as done by G02 with the {\it Chandra} data,
but we already know that the soft starburst contribution is not
significant  above 2 keV.}    we do not find  evidence of
variability:
 the $\chi^2$ test, applied to the binned
light curve ($\Delta T=5700$ sec), gives a probability of 50\%
that the data are consistent  with a constant flux during the
\emph{Beppo}SAX observation. It is worth noting  that this result
does not   exclude variability on different time scales and/or
irregular variability.

Finally, the XMM-{\it Newton} and \emph{Beppo}SAX observations are not simultaneous but
the MECS and XMM-{\it Newton} 2--10 keV fluxes, derived
assuming a simple power-law (PL) model,  are  comparable within the relative uncertainties; thus
a statistically significant flux difference is not found between the two data sets.

\section{Spectral analysis}

The XMM-{\it Newton} and \emph{Beppo}SAX data have been analyzed using standard
software packages  (FTOOLS 5.2, XSPEC 11.2). All  models discussed here have
been filtered for  the Galactic absorbing column density along the line of sight
($N_\mathrm{H} = 1.3\times 10^{20} $ cm$^{-2}$ Dickey \& Lockman, 1990). Unless
otherwise stated, errors will be given at the 90\% confidence level for one
interesting parameter ($\Delta\chi^2 = 2.71$) and  the line(s) energy will be
given in the source rest frame. The thermal emission  has been described  with
the MEKAL model in XSPEC  (Mewe et al. 1985). Abundances were kept solar;
keeping them free did not substantially change the results. Since MKN~231 does
not show significant flux variation we have first derived the cross-calibration
between MOS, pn and MECS and then, by assuming  that the cross-calibration
factor between MECS and PDS ranges from 0.7 to 0.95  (see Fiore, Guainazzi \&
Grandi 1999), we have tied together the XMM-{\it Newton} data and the PDS data.\\

From previous X-ray observations, performed with ASCA and  {\it
Chandra}, we know that the broad-band  (0.2--10 keV) spectrum of
MKN~231 appears to be more complex than a single power-law model.
In particular the {\it Chandra} data  presented by G02 show that
the circumnuclear starburst thermal emission,  which extends up
to $\sim 25''$ from the nucleus,  dominates the spectrum at $E <
2$ keV.  Due to the PSF of XMM-{\it Newton} and \emph{Beppo}SAX
detectors  the extracted counts in the low energy domain are
therefore ``contaminated" by the circumnuclear starburst. In contrast  
  the hard (E $>$ 2 keV) X-ray emission should be less
affected by the starburst  activity and more suitable for
investigating AGN related emission(s).  For these reasons we have
decided to focus our attention first on the 2--50 keV  spectrum,
and for simplicity (since we have enough statistics) to   use only
the pn and the PDS data.  In a second step we have also
considered the 0.5--2 keV source counts as well as  the data from all
the other detectors (MECS, MOS1 and MOS2).

\subsection{Fitting the Hard (2--50 keV) X-ray Energy Band: pn + PDS data}
 
In agreement with previous analysis  (Iwasasawa \cite{iwa99}, Turner 1999, MR00 and  G02)   when we fit
the   2--10 keV spectrum of MKN~231 with a single unabsorbed or absorbed PL model,   we find a  photon
index  ($\Gamma = 0.67\pm 0.11$ and  $\Gamma =  0.92_ {-0.23}^{+0.28}$, respectively)  which is 
unusually low  if compared with  the typical photon index of radio-quiet QSO ($\Gamma$ ranges from 1.7
to 2.3, Reeves  \& Turner \cite{reeves2000}; George et al. 2000). Furthermore the fit leaves positive
residuals around  6.4 keV,  strongly suggesting the presence of the iron emission line(s). The best fit
energy position of this line is  6.40$^{+0.15}_{-0.11}$ keV with an observed $EW = 300\pm 130$ eV, in
agreement with the ASCA data discussed by MR00 (in this first modelling the Fe line was constrained to
be unresolved). The flatness of  the observed X-ray spectrum is suggested  also by the high energy (E =
10--50 keV) PDS data: in fact  if we constrain the photon index to lie in the range 1.7-2.3, the PDS
data are underestimated by at least   a factor of 4. These results suggests: a) the presence of a 
heavily absorbed component and b) that below 10 keV  we are probably seeing only the reprocessed
radiation (through reflection and/or scattering, see Matt et al. 1991). In the following  we
investigate in turn two extreme possibilities: a reflection-dominated and a scattering-dominated
scenario.

\subsubsection{The reflection-dominated scenario}

In the  reflection-dominated scenario the  bulk of the 2--10 keV emission derives from
Compton reflection from optically thick material. When we consider reflection from
neutral matter a strong (i.e. EW as high as a few keV; Matt et al. 1996) Fe emission
line is also expected,  this line being produced by the same medium which reflects  the
primary continuum. The lack of a strong Fe emission line seems then to argue against a
reflection-dominated scenario. However,   the intensity and the energy centroid of the
Fe line depends on   the  ionization state of the medium (Matt et al. 1993, Ross et
al. 1999);  in particular, the intensity  decreases, compared  to  reflection from a
neutral medium, when the  ionization parameter  ($\xi$) is between 100 and 300,
due to the   resonant trapping opacity (Zycki  \& Czerny \cite{zyc}).

\vskip 0.5truecm

{\centerline {\it a) Neutral medium}}

\vskip 0.5truecm

A pure reflection-dominated continuum, where the reflecting medium
is neutral, cannot reproduce the 2--50 keV spectral properties of
MKN~231, leaving strong positive residuals above 7 keV.
Furthermore this model would require a  Fe emission line at 6.4
keV stronger than that measured. As previously noted  by MR00, a
prominent  scattered component  is needed to reproduce the
observed continuum and to dilute the Fe line.  We then tested the
pure reflection plus scattering  model using  a PL  plus  a pure
reflected  continuum (PEXRAV\footnote{ Given  the present
statistics above 20 keV  we fixed $cos(i)=45$, $Z=Z_\odot$ and
$E_c=200$ keV and allowed  only the photon index and the
normalization to vary.} model in XSPEC; Magdziarz \& Zdziarski 1995)  
plus  an unresolved  gaussian line. The best fit photon
index ($\Gamma\sim 1.3$) is still low if compared with  the
typical photon index of radio-quiet QSOs. Constraining the   photon
index to lie in the range 1.7--2.3 we cannot find an acceptable
fit. The results  are summarized in Table 1, model A and model B. These
models do not consistently explain the low observed
EW of the line and the flatness of the spectrum for  an intrinsic standard AGN emission. \\

\vskip 0.5truecm

{\centerline {\it b) Ionized medium}}

\vskip 0.5truecm

An alternative explanation for the observed low  EW   of the Fe line 
   invokes the ionization state of the reflecting medium.

Since the reflected continuum from mildly  ionized material predicts
the presence of  structured features in the low-energy domain (see e.g.
Ross \& Fabian. 1993) we are forced to consider  also the pn counts below 2
keV.  On the other hand, below 2 keV we also have to  take into account
the contribution from the circumnuclear starburst.

Thus, we have  first tested a 3-component model composed of: a)  
thermal emission to take into account the starburst contribution
at low energies; b) a pure reflected component from mildly ionized
material (PEXRIV\footnote{Given  the present   statistics  above
20 keV we fixed $cos(i)=45$, $Z=Z_\odot$ and $E_c=200$ keV and
allowed only the photon index and the normalization to vary. The ionization parameter $\xi$
was constrained to lie  in the range 100-300.} model in XSPEC;
Magdziarz \& Zdziarski 1995)
 and  c) a gaussian line.

This model gives a poor representation of the broad band  (pn +
PDS data) properties of MKN 231  ($\chi^2/\nu=102.6/65$; see Table
1 model C), leaving strong positive residuals at energies above 7
keV and underestimating by a factor 5 the counts in the PDS energy
range; all these features strongly suggest  the presence of a
heavily absorbed  ($N_{\rm H} \gae $ $10^{24}$ cm$^{-2}$) component. We
have thus added a highly absorbed PL component  (described by the
PLCABS\footnote{Given that the intrinsic $N_{\rm H}$ is expected to be
high ($\sim 10^{24}$cm$^{-2}$) we have used this model, which
takes into account also the Compton scattering.} model in XSPEC;
Yaqoob 1997) whose photon index has been tied to the photon index
of the reflected component from ionized material.

This model gives a better broad band description  of the spectrum of MKN~231 (see Table 1 model D), but it  requires a very
 steep intrinsic photon index ($\Gamma = 2.95\pm 0.07$).

 Finally,  given that MKN~231 has a strong circumnuclear starburst  (see Introduction), and
given the  size of the source counts extraction radius,  we expect a contribution from X-ray
binaries, and in particular from High-Mass X-ray binaries (Franceschini et al., 2003; Persic et
al., in preparation). This latter component has been described assuming an absorbed cutoff-power
law model of the  form $E^{-\Gamma} exp^{-h\nu/kT}$ with photon index $\Gamma \sim 1.1$ and cutoff
energy $kT \sim 10$ keV (cfr. Persic \& Raphaeli 2002).

This final model gives a good description of the X-ray spectra of MKN 231. In
Fig.~1 (left panel) we  show the data and  the residuals  while the best fit spectral
parameters are reported in Table 2 (Model A).

The  line energy position  is $6.38\pm 0.19$     at 90\%
confidence level, in  agreement with the value (6.5 keV) expected for this range of
ionization of the mirror. The EW of the line is $\sim 200 $ eV with respect to
the reflected component, which is consistent with being produced by  a mildly
ionized mirror (Ross et al. 1999).   By allowing the line width to vary we
obtain $ \sigma\sim 0.1$ keV, consistent  with the XMM-{\it Newton} spectral
resolution. Thus at the  spectral resolution of XMM-{\it Newton} the detected Fe
line   is unresolved.  No statistical difference  for the  best fit
parameters reported in Table 2 is found  if the line width is allowed    to vary.

In conclusion, a pure reflection-dominated continuum from neutral
material cannot consistently explain the low Fe
emission line EW   and the flatness of the spectrum. A viable solution
is the pure reflection continuum from slightly ionized material
combined with a  highly  absorbed component. Adding to this ``AGN
baseline model" the spectral components  which are likely
associated to the starburst (a soft thermal component and  a PL
component associated to the High Mass X-ray Binaries) we found a
value of $\sim 2.5$ for the AGN intrinsic photon index.  This steep intrinsic
photon index is not so striking if we consider that MKN~231 shares
many spectral characteristic of NLS1s (i.e. weak [O$_{III}$],
relatively narrow Balmer emission lines and strong optical
Fe$_{II}$ emission;  Mathur 2000; Brandt \& Gallagher
\cite{brandt}). According to this  model, hereafter  called the
``AGN reflection-dominated" model, the intrinsic 2--10 keV
luminosity  of the AGN powering MKN 231 is L$_{(2-10)} \sim
2\times 10^{44}$ ergs s$^{-1}$,   in agreement with the
value predicted by MR00  based on ASCA data.

\begin{figure*}
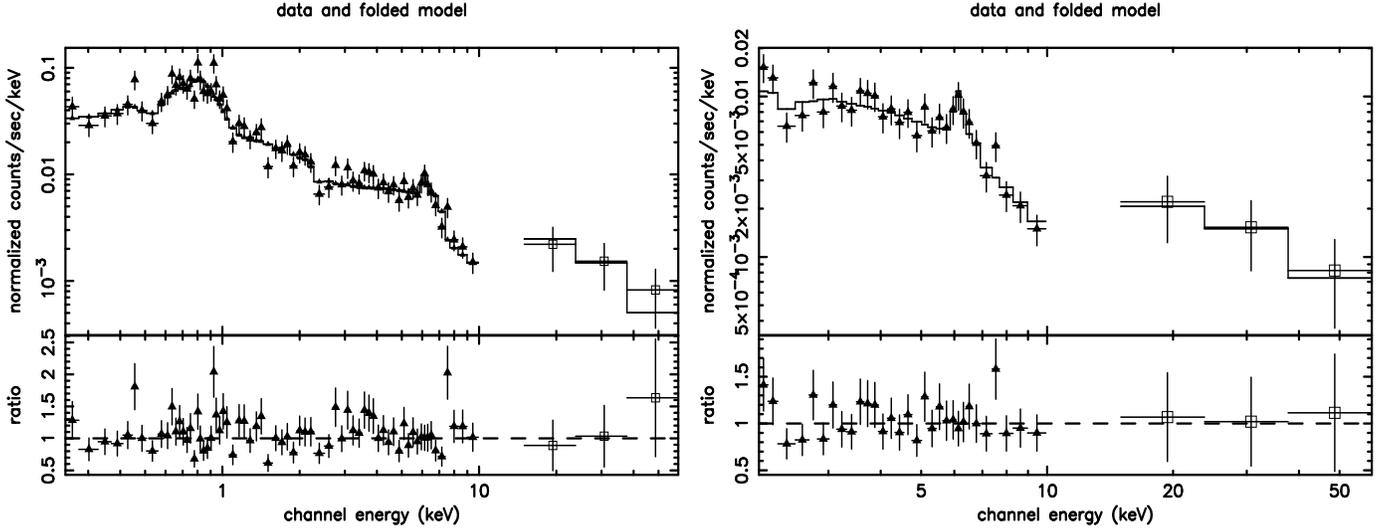

\begin{tabular}{cc}
\hskip-0.1truecm\psfig{file=fig1a.ps,height=9.0cm,width=7cm,angle=-90}&
\hskip-0.2 truecm\psfig{file=fig1b.ps ,height=9.0cm,width=7cm,angle=-90}\\

\end{tabular}
\caption{Left panel: 0.2--50 keV pn (filled triangles) plus PDS (empty squares) data and ratio between data
and the best fit reflection   model (model A in Table~2). Right panel: 2--50 keV pn (filled triangles) plus PDS (empty squares) data and ratio between data
and the  scattering dominated model discussed in Sect.~4.1.2}

\end{figure*}

\begin{table*}[htb]
\caption{Pn plus PDS data: rest-frame results from the  basic
reflection-dominated  models discussed in Sect. 4.1.1}

   \begin{tabular}[h]{l ll| ll}
      \hline
      \hline

     \multicolumn{3}{c|}{AGN Continuum} &\multicolumn{2}{c}{Fe line} \\

 MODEL & Parameter & Value& Parameter & Value\\
\hline\hline
    	& $\Gamma^a$ & $2.65^{+0.19}_{-0.19}$     & $E_{\mathrm{k}\alpha}$    & $6.41^{+0.22}_{-0.18}$ keV\\
     	& $N_H^b$	& / &                        $EW$ & $<0.24$ keV        \\
   A) refl PL+Fe line & R  &  $1.3^{*}_{-0.1}$       &        &             \\

    &$N_H^c$&     /&   $\chi^2/dof$    &      35.61/29     \\

    &&              &&  \\

 \hline
 \hline
                            &&&&\\
   	& $\Gamma^a$ & $1.32^{+0.37}_{-0.59}$     & $E_{\mathrm{k}\alpha}$    &$6.42^{+0.22}_{-0.16}$ keV\\
    	& $N_H^b$	& / &                         $EW$ &   $0.28^{+0.30}_{-0.20}$ keV     \\

      B) refl PL+Fe line +scatt PL & R  &    $1.01^{+5400}_{-1.01}$  & $\sigma$     &  $   0.16^{+0.30}_{-0.16} $ keV    \\

    &$N_H^c$& $2.14^{+1.35}_{-1.25}$ &   $\chi^2/dof$    &   19.6/26    \\

    &&              &&  \\

  \hline
\hline
                            &&&&\\
   	& $\Gamma^a$ &$2.84^{+0.13}_{-0.12}$       & $E_{\mathrm{k}\alpha}$    &$6.40^{+0.24}_{-0.19}$ keV \\
    	& $N_H^b$	& 0.19$^{+0.11}_{-0.10}$ &                       $EW$ &     $0.18^{+0.18}_{-0.14}$    keV  \\
 
 C)ioniz refl+Fe line        & R  & $0.94^{+1150}_{-0.16}$         &   $\sigma $       &   $   0.10^{+0.26}_{-0.10} $ keV          \\

    &$N_H^c$& / &   $\chi^2/dof$    &  102.6/65     \\

    &&              &&  \\
    
  \hline  
  \hline
                            &&&&\\
   	& $\Gamma^a$ &$2.95^{+0.15}_{-0.16}$       & $E_{\mathrm{k}\alpha}$    &  $6.39^{+0.18}_{-0.20}$ keV\\
    	& $N_H^b$	& 0.24$^{+0.15}_{-0.10}$ &                         $EW$ &     $0.18^{+0.17}_{-0.15}$    keV  \\
   D)ioniz refl+Fe line+trans PL      & R  &  $1.04^{+1042}_{-0.15}$         &    $\sigma $      &   $   0.10^{+0.25}_{-0.10} $ keV          \\

    &$N_H^c$&  $272.6^{+99.9}_{-68.0}$ &   $\chi^2/dof$    &    89.0/61    \\

    &&              &&  \\
  \hline  
  
\end{tabular}
\\ 

For model A and model B we have considered the 2--50 keV data, while for model C and D we have considered the  0.2--50 keV data. For model C
and D  the ionization parameter has been constrained to lie within the range 100-300 erg cm/s.\\ 
$^{a}$ intrinsic photon index of  the reflected and scattered component; in model D the photon indices of the
transmitted and reflected component have been tied together; \\
$^{b}$ absorption of the reflected component in units of $10^{22}$cm$^{-2}$;\\
$^{c}$ absorption of the scattered (transmitted) component in units of $10^{22}$cm$^{-2}$;\\
$^{*}$ unconstrained parameter.\\


\end{table*}

\subsubsection {The scattering-dominated scenario}

Another possible explanation for the   hard MKN 231 X-ray continuum  
 flatness  and the lack of a strong Fe emission line is  a scattering-dominated
scenario in which  we see the AGN mainly  through  scattered emission (see G02
modeling of the {\it Chandra} data). In this context we assume that the Fe
emission line is produced by transmission and is diluted from the scattered
component.

To test this scenario we have used a 3-component model composed of:
a) a highly absorbed PL  representing
the intrinsic AGN emission transmitted through the Compton-thick screen;
b) a PL having the same photon index as the transmitted component, representing
the scattered AGN emission and
c) a gaussian line.
This model gives a fairly good description of the data ($\chi^2/\nu=20.47/27$)
and, as shown in Fig.~1 (right panel), no strong residuals are present.

The intrinsic AGN emission is described by a PL model having a photon index of
$\Gamma=1.69^{+0.49}_{-0.70}$ which is filtered by an intrinsic absorption
column  density of $N_{\rm H} =  1.4^{+0.9}_{-0.5}\times 10^{24}$cm$^{-2}$. The 
energy position of the Gaussian line best fit  value is $E =
6.39^{+0.13}_{-0.12}$ keV,  which is consistent with the neutral Fe-K$\alpha$
line.  Allowing the line width to vary we obtain $\sigma=0.2\pm0.18\ $keV   (we note however that if we
allow  the line width to vary the fit does not improve significantly). 
The  Fe line EW    is $\sim$ 260 eV with respect to the observed continuum
and $\sim 1.2$ keV with respect to the absorbed PL.  This latter value is
consistent with that expected to be produced by transmission in the same medium
that absorbs the primary nuclear emission (Leahy \& Creighton 1993). The
scattered component is also absorbed by an intrinsic absorption column  density
of $N_{\rm H} =  4.1^{+1.8}_{-3.2}\times 10^{22}$cm$^{-2}$.

In the 2--10 keV band the observed flux  is $7\times 10^{-13}$ \ecs   and  the intrinsic
2--10 keV luminosity, de-absorbed for the intrinsic $N_{\rm H}$, is L$_{(2-10)}=4\times
10^{43}$ ergs s$^{-1}$.

In summary this model provides another possible explanation for   the
flat MKN 231 observed  2--50 keV  continuum   and its line properties.

Furthermore,
although not well constrained, the  best fit photon index is now fully
consistent with the range usually found in AGN ($\Gamma \simeq 1.7-2.3$).
In the following we will refer to  this model as the ``AGN scattering-dominated"
model.

\begin{table*}[htb]
\caption{Broad-band best fit models and  parameters: Model A refers to the
reflection-dominated model, while model B refers to the scattering-dominated model. Both  models include the starburst components.}

   \begin{tabular}[h]{c ll| ll}
      \hline
      \hline

     \multicolumn{3}{c|}{Starburst } &\multicolumn{2}{c}{AGN} \\

  & Parameter & Value& Parameter & Value\\
\hline\hline
    & kT  & $0.66^{+0.05}_{-0.05}$ keV       &   $\Gamma$ &$2.48^{+0.20}_{-0.11}$ \\
     & / &                       &   $N_H^b$ & $0.20^{+0.08}_{-0.04}$       \\
 MODEL A  & $\Gamma_{BIN}$ &$1.1^\star$          &    $N_H^c$& $264.6^{+173.2}_{-85.1}$             \\
    & $N_H^a$ & $4.22^{+16.9}_{-4.22} $ &         $E_{\mathrm{k}\alpha}$ & $6.38^{+0.19}_{-0.17}$ keV       \\
    &&              &EW & $0.20^{+0.16}_{-0.16}$ keV    \\

    &&              &&  \\

 \hline
 \hline
                            &&&&\\
    &kT$_1$  &$0.35^{+0.09}_{-0.05}$ keV& $\Gamma$ &$1.83^{+0.12}_{-0.69}$         \\
    &kT$_2$ & $0.91^{+0.15}_{-0.10}$ keV& $N_H^b$ &$12.16^{+2.67}_{-2.03}$ \\
 MODEL B& $\Gamma_{BIN}$ &$1.1^{\star}$     & $N_H^c$&$177.5^{+105.8}_{-110.0}$\\
    & $N_H^a$ &$1.7^{+6.4}_{-1.7}$   &$E_{\mathrm{k}\alpha}$  & $6.39^{+0.15}_{-0.14}$ keV\\
    &&              &EW & $0.29^{+0.18}_{-0.18}$ keV    \\
   \hline

\end{tabular}
\\

$^{a}$ absorption of the HMXBs component in units of $10^{20}$\cm2\\
$^{b}$ absorption of the reflected/scattered component (model A/model B) in units of $10^{22}$\cm2 \\
$^{c}$ absorption of the transmitted component in units of $10^{22}$\cm2;\\
the symbol $^{\star}$ indicate that the parameter has been kept fixed.
\end{table*}

\begin{table*}[t!]
\caption{Observed X-ray Fluxes and intrinsic X-ray luminosities derived in the two competing scenarios }

   \begin{tabular}[h]{cccc ccc ccc }
      \hline

  &  \multicolumn{3}{c}{Flux (0.5--10) keV} & & \multicolumn{2}{c}{L(0.5--2) keV} &&\multicolumn{2}{c}{L(2--10) keV} \\
  &  \multicolumn{3}{c}{10$^{-13}$ ergs cm$^{-2}$ s$^{-1}$}   & &\multicolumn{2}{c}{10$^{43}$ ergs s$^{-1}$}&   &\multicolumn{2}{c}{10$^{43}$ ergs s$^{-1}$} \\
   &  &  &      &   &  &  &  && \\
 \cline{2-4}\cline{6-7}\cline{9-10}
    &  &  &      &   &  &  &  && \\

MODEL &  TOTAL           & AGN&  Starburst &              & AGN&Starburst&& AGN&Starburst\\
\hline\hline

  &  &  &      &   &  &  &  && \\

A  & 7.9 & 6.3 & 1.6     &    & 33 & 0.06 &  &20&0.07 \\
 &  &  &      &   &  &  &  && \\

B & 8.2 &  5.0 & 3.2      &   &2.9  &0.09   &  &4.5&0.16 \\
 \hline
  \end{tabular}

Model A refers to reflection-dominated model\\
Model B refers to scattering-dominated model\\
\end{table*}

\subsection {The broad band (0.5--50 keV) X-ray spectral properties of MKN 231}

In the previous section, using the pn and the PDS data, we   found that two
competitive models (reflection-dominated and scattering-dominated) can describe
the MKN~231  AGN  emission.  In this section  these two  models will be  
used as   starting points to study the broad band (0.5--50 keV) X-ray spectral
properties of MKN~231.  Low energy data ($E < 2$ keV)  and  all   the 
detectors MOS1, MOS2, pn, MECS together with the  PDS data will be used. \\

In the  ionized reflection model  case we found that the model derived for the
pn and the PDS data (which already include the spectral components expected
from the starburst, see Sect.~4.1.1) is able to reproduce all  data with no
strong differences for the previously determined best-fit parameters. In
Fig.~2  (left panels)  we show the model and the residuals  for all the
detectors, while the best fit spectral parameters are reported in Table~2
(Model A).  Within the framework of this modelling the intrinsic luminosity of
the AGN powering MKN~231 is: L$_{(0.5-2)}=3.3\times 10^{44}$ ergs s$^{-1}$ and 
L$_{(2-10)}=2\times 10^{44}$ ergs s$^{-1}$ (see Table~3).\\

In  the scattering-dominated scenario   case the
residuals below 2 keV are indicative of  thermal emission.
Indeed, two thermal   components are needed to reproduce the
soft X-ray emission of MKN 231 ($kT \sim 0.35^{+0.09}_{-0.05}$ keV
and  $kT \sim 0.91^{+0.15}_{-0.10}$ keV). For the  HMXBs  emission
we have added an
absorbed cutoff PL component,   following what already done in Sect.~4.1.1;   moreover,  this component
is needed to  reproduce the     continuum   detected in the 1--3 keV range. The best-fit spectral parameters
derived with this broad-band, multi-component  modelling are
reported in Table~2 (Model B; $\chi^2/\nu =212.9/187$). Note that  the best-fit photon index for the AGN
component
 is now  $\Gamma=1.83^{+0.12}_{-0.69}$. The derived intrinsic 2--10 keV
luminosity for the AGN is    L$_{(2-10)}=4.5 \times 10^{43}$ ergs s$^{-1}$
A summary view of the model is reported in
Fig.~2 (right panel).\\

\begin{figure*}
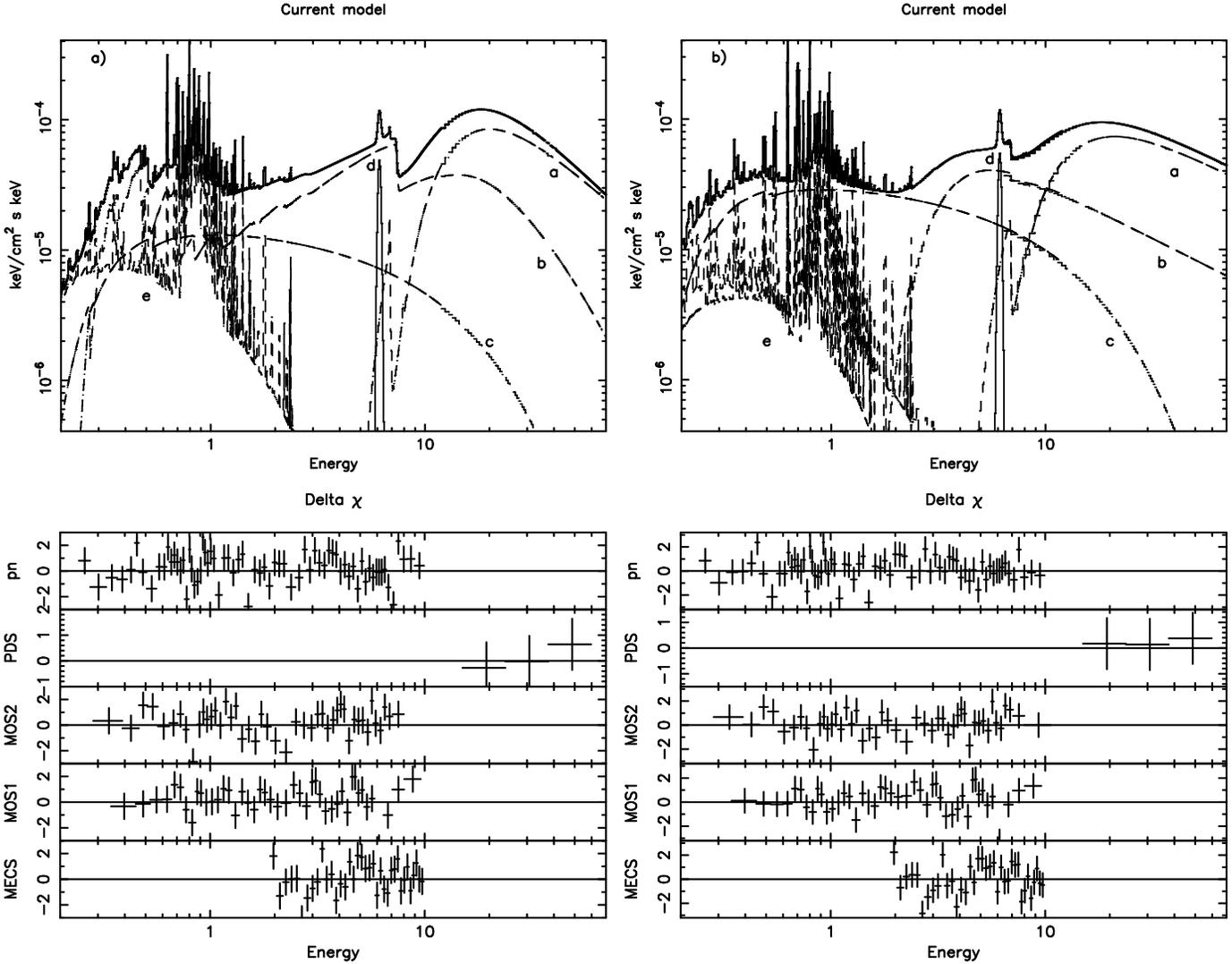

\begin{tabular}{cc}
\hskip-0.5truecm\psfig{file=fig2a.ps,height=9.0cm,width=7cm,angle=-90}&
\hskip-0.2truecm\psfig{file=fig2b.ps,height=9.0cm,width=7cm,angle=-90}\\
\hskip-0.5truecm\psfig{file=fig2c.ps,height=9.0cm,width=7cm,angle=-90 }&
\hskip-0.2truecm\psfig{file=fig2d.ps,height=9.0cm,width=7cm,angle=-90 }\\

\end{tabular}
\caption{0.2--50 keV data coverage.
 Upper panels:  best-fit models for the
reflected-dominated (left panel) and scattering-dominated scenario (right panel). The spectral components are: a) a highly absorbed
PL AGN component;  b)left panel: a pure reflection AGN component from slightly ionized
material or right panel: scattered AGN component; c) a cutoff PL  component associated with the binaries in the
starburst; d) a narrow Gaussian line at 6.39 keV;
 and e) a thermal emission component associated with the starburst. Lower
panels:  residuals for the different detectors (from up to down: pn, PDS, MOS2, MOS1, MECS).}

\end{figure*}

\section{Discussion}

\subsection{The AGN component in MKN~231}

The AGN spectral behavior can be described with two models which differ mainly
at  E $\lae$ 8 keV. Both these models require   a heavily absorbed
component to account for the  X-ray emission  above 10 keV detected with the
PDS. The photon indices derived with these models, although not well
constrained, are consistent with the typical values of AGNs. Furthermore the PDS
data have  proven,  for the first time, the presence of a thick column density
screen ($N_{\rm H} \sim 2\times 10^{24}$ cm$^{-2}$) which blocks the primary X-ray
emission. The 2--10 keV observed spectrum  could be the result of    reflection
(reflection-dominated scenario) or of  scattering  (scattering-dominated
scenario) which emerges from the Compton-thick screen. Clearly the real
physical situation could be an intermediate one where we see both components.
Since the complexity of the 2--10 keV continuum prevents us from
distinguishing, on a statistical basis, between these two models, we have checked
if the physical parameters derived in both scenarios are in agreement with what
is known about ULIRGs, starburst galaxies and in particular with the broad-band
properties of MKN~231.\\

\begin{figure*}[t]
\begin{tabular}{cc}
\hskip-0.5truecm\psfig{file=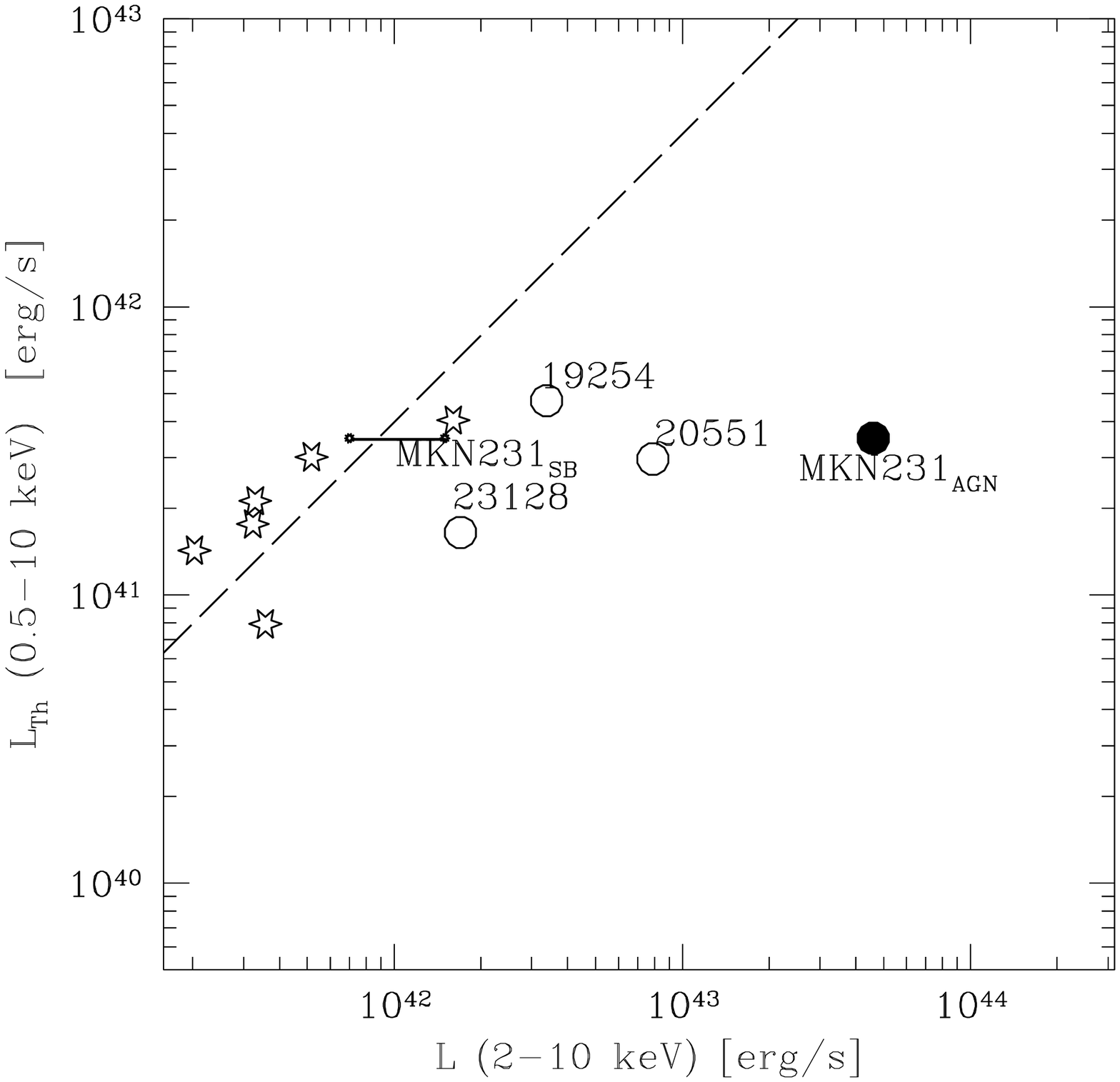,height=8.0cm,width=8cm}&
 
 \hskip-0.2truecm\psfig{file=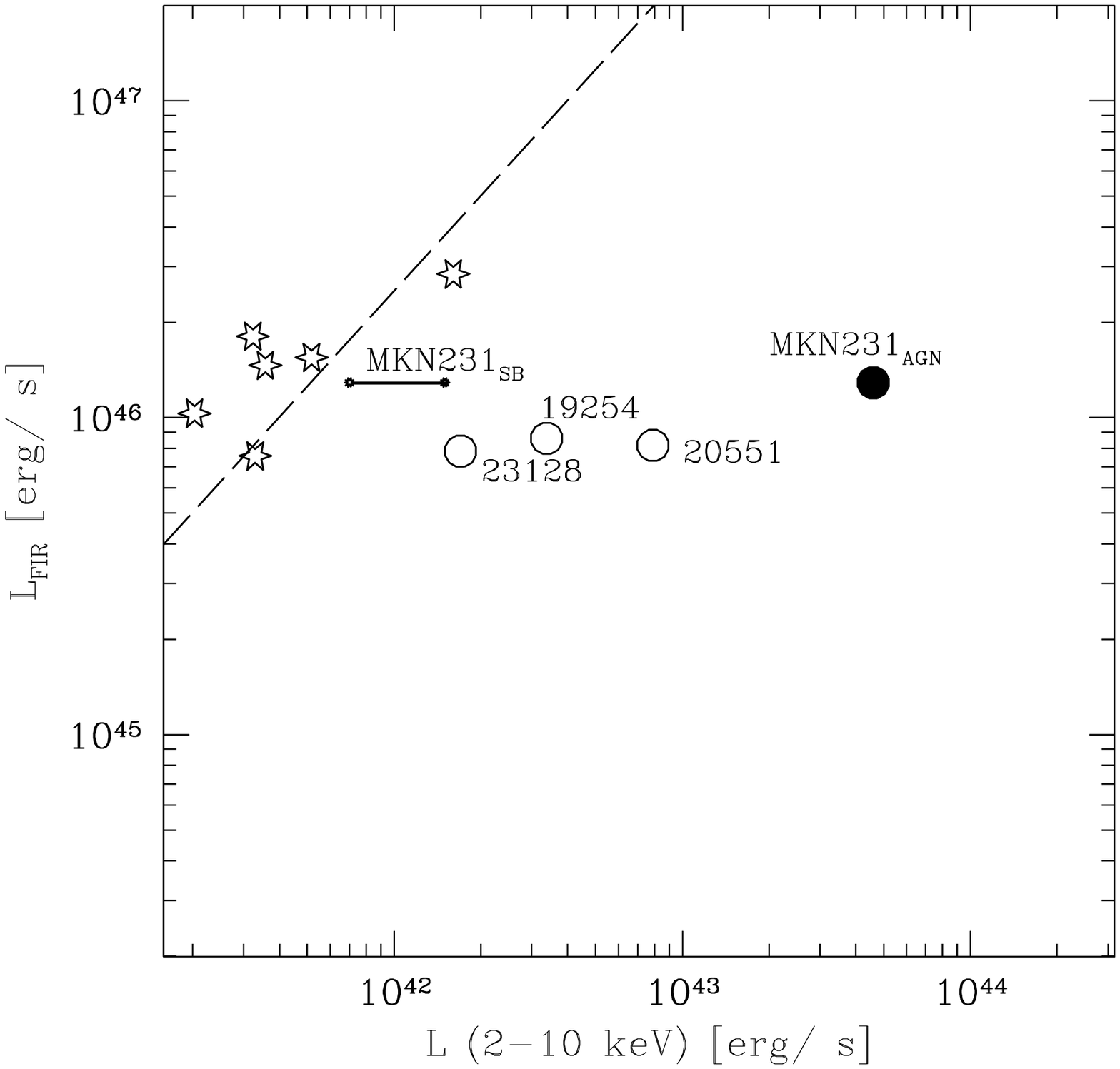,height=8.0cm,width=8cm}
 \\
\end{tabular}

\caption{Luminosity-luminosity plots of various emission components identified in the XMM-Newton and \emph{Beppo}SAX spectrum
of MKN~231  compared with  the ULIRGs in Franceschini et al. (2003).  Starred symbols refer to sources classified  as dominated
by starburst, open circles to AGN-dominated sources. For MKN~231 we report (black solid line) the range of the hard X-ray
luminosity of the  binary components and the lower limit of the AGN (filled circle) hard X-ray
luminosity;  we have also assumed that the starburst activity in MKN~231 contributes to 50\% of  $L_{FIR}$. Left panel : luminosity-luminosity plot of the thermal versus PL components.
Right panels: $L_{FIR}$  against the luminosities of the PL components. Dashed
lines indicate proportionality relations. } \label{fig_comp} \end{figure*}

We identify the optically thick absorber with the BAL outflows
which originate close to the central source.   The ionized
matter, which scatters or reflects the primary emission, could be
the inner part of this  outflow, or ``shielding gas''  (Murray et
al. 1995) in agreement with the scenario proposed by G02. In both
models proposed to take into account  the AGN emission the scattered/reflected
components are absorbed. This latter absorbing medium could be
identified with the starburst  regions (see e.g. Levenson et al.
2001)  or with a different line of sight through the BAL wind. \\
Beside the flat shape of  the observed MKN~231 X-ray continuum,
the other topic that emerged, since the first hard X-ray
observations, is the low 2--10 keV luminosity.  If we assume that
the far-infrared emission of MKN~231 is mainly powered by  AGN
activity then the intrinsic  X-ray luminosity should be $\sim
10^{45}$ ergs s$^{-1}$ (Risaliti et al. 2000).  In both models the intrinsic
2--10 keV X-ray luminosity can be measured from the absorbed
component and is $\sim 1^{+1.0}_{-0.5} \times 10^{44}$ erg s$^{-1}$. 
Although the balance between the X-ray and bolometric luminosity
is derived assuming a standard type 1 AGN, this  estimate   
suggests that the contribution
of the starburst activity  to the FIR luminosity could be significantly higher than that of the AGN.\\

\subsection{The starburst component in MKN~231}

In both  AGN models proposed  we found  a strong  soft X-ray component which we
describe with thermal emission(s) associated to the starburst activity. The derived temperature(s)  are in agreement
with the results obtained  for the extended emission  by G02 using {\it Chandra }
data ($kT\sim 0.3 $ keV and $kT\sim 1.1$ keV)
and consistent with  that
measured in other well known bona-fide luminous starburst galaxies.\\
The energetics of this thermal component(s)
(L$_{(0.5-2)}=6-9\times 10^{41}$ erg s$^{-1}$ depending on the AGN
modelling) is comparable with what has been previously found for
ULIRGs (Franceschini et al. 2003, Ptak et al. 2003). 
The measured luminosities are  in agreement with the one estimated
using ROSAT and ASCA data (Iwasawa 1999; Turner 1999), but about a
factor 4-6  higher than that found by G02 using {\it Chandra} data
($L_{(0.5-2)}= 10^{41.2}$ erg s$^{-1}$).

For the hard X-ray emission, which we attribute to the HMXB,  we found that 
L$_{(2-10)}$ ranges from $6.9\times 10^{41}$ erg s$^{-1}$ to $1.6\times
10^{42}$ erg s$^{-1}$ (in the reflection and in the scattering-dominated
scenario respectively).  This spectral component is expected in a source like
MKN~231 with a strong nuclear starburst  and has to be taken into account in
the spectral modelling. We have  checked if the X-ray luminosity of the cutoff
PL component is in agreement with the results obtained by Franceschini et al.
(2003)  for a sample of ULIRGs observed with XMM-{\it Newton}. These
observations have shown that   for all the starburst  dominated  ULIRGs  the
hard X-ray luminosity of the X-ray binaries  clearly correlates with the FIR
luminosity.

In Fig.~3 we show the result of the comparisons  between the X-ray
emission due to  starburst and to   AGN activity  and
$L_{FIR}$ for the 10 ULIRGs observed  with XMM-{\it Newton}; in
these plots  we have assumed that  MKN~231  starburst
activity contributes for $\sim 50$\% to  the $L_{FIR}$.

The range of the 2--10 keV  HMXB   luminosity derived with the two different AGN models,
clearly  locates MKN~231 in the region occupied by the starburst-dominated ULIRGs.  This
result  gives confidence that the  power-law component originates from  the population of
HMXB  expected to be present in the circumnuclear starburst region of MKN~231. \\ We  have
also compared the    star-formation rate (SFR) predicted  from the 2--10 X-ray emission 
with that derived by the FIR luminosity. To do that we have  taken into account that  the
latter contributes to the  2--10 keV emission   mainly through the HMXB emission in case of
intense and recent star-forming activity (Franceschini et al. 2003, Persic et al. in prep.)\footnote{The relation  between
HMXB luminosity and SFR has been recently discussed in the literature (i.e. Grimm
et al. 2003; Gilfanov et al. 2004 and Ranalli et al. 2003).}.  This assumption
is consistent  with a significant contribution of the starburst activity to 
the $L_{FIR}$. Our result is in fair agreement with the results  obtained with
the modelling of the 1--1000 $\mu m$ Spectral Energy Distribution by Farrah et
al. (2003).

 Considering the heavily obscured PL detected  with the PDS and
the above starburst scenario MKN~231 no longer appears as an
unusually X-ray faint AGN.
In conclusion,   XMM-{\it Newton} and \emph{Beppo}SAX observations suggest   
that starburst
activity   strongly  contributes  to the $L_{FIR}$ of MKN~231, 
in agreement with  multiwavelength  analysis. \\

\section{Summary}

By combining the XMM-Newton spectral resolution and throughput  with the high-energy bandpass of BeppoSAX we have investigated in deeper detail than
previously  possible the broad-band (0.2--50 keV)  X-ray spectral properties of
MKN 231.\\
We confirm that  MKN~231 is  a complex source whose  X-ray spectrum has clear
signatures of both  a highly obscured AGN and strong starburst activity.  \\
The detection, in the PDS energy range, of a highly absorbed ($N_{\rm H}\sim 
2\times 10 ^{24}$ cm$^{-2}$) component  for the first  time makes it possible to measure the
AGN intrinsic 2--10 keV luminosity of MKN 231.  This luminosity, which ranges between 
$4.5\times 10^{43}$ ergs s$^{-1}$  and $   2\times 10^{44}$ ergs s$^{-1}$, is at least 
an order of magnitude higher  than what has been previously measured.\\
We found that the observed flat 2--10 keV X-ray continuum of  MKN~231 can 
be explained assuming that below 10 keV we are seeing only reprocessed 
radiation through reflection and/or scattering.\\
The starburst activity significantly contributes to the soft (E $<$ 2 keV) 
X-ray emission ($L_{\rm {Th\; (0.5-2 keV)}}\sim 3 \times 10^{41}$ergs s$^{-1}$), although a relevant contribution due 
to HMXB is also measured in the 2--10 keV energy range ($L_{\rm {HMXB\; (2-10 keV)}}= 0.7-1.6 \times 10^{42}$ergs s$^{-1}$).

 \begin{acknowledgements}  Based on observations obtained with XMM-Newton, an ESA science mission with instruments and 
 contributions directly founded by ESA member states and USA (NASA).
 This research has made use of SAXDAS linearized  and cleaned event files produced at the BeppoSAX Science Data Center,   and the NASA/IPAC extragalactic database (NED).
 We thank all the members working at the above mentioned projects.  We thank R. Saxton for his suggestions during the XMM-{\it Newton} data
  processing, and we thank L. Ballo, A. Caccianiga and the anonymous referees for useful
comments. 
   This work has received financial support from
ASI (I/R/037/01)  and from ASI (I/R/062/02) under the project ``Cosmologia Osservativa con
XMM-Newton". 
\end{acknowledgements}

\end{document}